\documentclass[aps,prd,preprintnumbers,groupedaddress,nofootinbib,showpacs]{revtex4}

\usepackage{graphicx}
\usepackage{latexsym}
\usepackage{amsfonts}
\usepackage{amssymb}
\usepackage{amsmath}
\usepackage{slashed}
\usepackage{array}
\usepackage{feynmp}
\usepackage{hyperref}
\usepackage{url}

\newcommand{\beq}{\begin{equation}}
\newcommand{\eeq}{\end{equation}}
\newcommand{\beqs}{\begin{eqnarray}}
\newcommand{\eeqs}{\end{eqnarray}}

\begin{document}
\title{Thermal Dark Matter from a Confining Sector}

\author{Matthew R.~Buckley$^{1}$ and Ethan T.~Neil$^2$}
\affiliation{$^1$Center for Particle Astrophysics, Fermi National Accelerator Laboratory, Batavia, IL 60510, USA}
\affiliation{$^2$Theoretical Physics Department, Fermi National Accelerator Laboratory, Batavia, IL 60510, USA}
\preprint{FERMILAB-PUB-12-533-A-T}
\date{\today}

\begin{abstract}
We study a class of dark matter models in which the dark matter is a baryon-like composite particle of a confining 
gauge group and also a pseudo-Nambu-Goldstone boson associated with the breaking of an enhanced chiral symmetry group.  The approximate symmetry decouples the dark matter mass from the confinement scale of the new gauge group, leading to correct thermal relic abundances for dark matter masses far below the unitary bound, avoiding the typical conclusion of thermally produced composite dark matter.  We explore the available parameter space in a minimal example model based on an SU$(2)$ gauge group, and discuss prospects for experimental detection.
\end{abstract}


\maketitle

\section{Introduction \label{sec:intro}}

The gravitational footprint of dark matter in the Universe provides irrefutable evidence of the existence of physics beyond the Standard Model. This new physics comes in the form of a new massive particle with no electromagnetic or strong force interactions, composing $\sim 25\%$ of the Universe's matter density. Despite decades of experimental work, no unambiguous direct evidence of the nature of this new particle has been found.  

The leading class of theoretical explanations assumes that the dark matter particle is a thermal relic of the early Universe, with a present-day abundance set by the pair annihilation cross section into Standard Model particles. The interest in this solution is largely motivated by an intriguing coincidence: a dark matter candidate with approximately weak scale masses and couplings (a Weakly Interacting Massive Particle, or WIMP) would naturally freeze out with the correct relic density. 

This ``WIMP Miracle'' has received even more theoretical and experimental attention due to the presence of such particles in many of the solutions to the naturalness and hierarchy problems of electroweak symmetry breaking. Of these, the best known is the neutralino in supersymmetric models. However, it should be noted that any particle with the appropriate ratio of mass to cross-section can provide a good thermal dark matter candidate. Such models are sometimes known as WIMPless \cite{Feng:2008ya,Feng:2008mu}.

Dark matter as a thermal relic is of course not the only possible scenario. It could be a non-thermally produced axion~\cite{Abbott:1982af,Dine:1982ah,Preskill:1982cy}. Or dark matter might, like baryons, possess an inherent asymmetry \cite{Nussinov:1985xr,Barr:1990ca,Barr:1991qn,Dodelson:1991iv,Kaplan:1991ah,Thomas:1995ze,Kuzmin:1996he,Fujii:2002aj,Gudnason:2006ug,Kaplan:2009ag,An:2009vq,Cohen:2009fz,Cohen:2010kn,Davoudiasl:2010am,Hall:2010jx,Kribs:2009fy,Shelton:2010ta,Blennow:2010qp,Buckley:2010ui,Buckley:2011kk,DelNobile:2011je,Graesser:2011wi,Gu:2010ft,Haba:2010bm,McDonald:2011zz,Blum:2012nf,Buckley:2011ye,Cui:2011ab,Davoudiasl:2012uw,Tulin:2012re, Gu:2012fg}. In this case, the two asymmetries could be related, either by high-dimension interactions that violate both baryon and dark matter numbers, or through non-perturbative effects, such as the $SU(2)_L$ sphaleron. However, such asymmetric models require an annihilation cross section in the early Universe at least as large as that of a thermal candidate \cite{Buckley:2011kk}.

In this paper, we introduce a new candidate for thermally produced dark matter, in which the dark matter particle is a composite stable pseudo-Nambu-Goldstone boson built of fundamental fermions bound together by a confining gauge force, which we call ``ectocolor.''\footnote{{\it ecto-} outside (Latin) } We will assume the fundamental fermion mass is much less than the confinement scale (reversing this inequality leads to a class of models known as ``thetons" \cite{Khlopov:1980ar} or ``quirks'' \cite{Kang:2008ea,Kribs:2009fy,Harnik:2011mv}). The key feature of our model is the requirement that the fermions are charged under a real or pseudo-real representation of the ectocolor gauge group.  The canonical example is the ${\bf 2}$ representation of $SU(2)$, which we will use throughout this paper as an explicit realization.

We note that the possibility of using stable pseudo-Nambu-Goldstone bosons as a thermal dark matter candidate has been considered previously in the context of partially-gauged technicolor \cite{Belyaev:2010kp} and little Higgs models \cite{Frigerio:2012uc}.  In these contexts, the direct connection to electroweak symmetry breaking gives additional motivation for the strongly-coupled sector which gives rise to the dark matter, but it also leads to significant constraints on the possible parameter space.  In particular, thermal production with the correct relic abundance is only found to be possible for a relatively narrow range of PNGB dark matter masses near the electroweak scale.  Here we consider the dark matter sector independent of electroweak symmetry breaking, leading to a larger viable parameter space.

By restricting ourselves to real or pseudo-real representations, the spectrum of light pseudo-Nambu-Goldstone bosons can contain both unstable mesons and stable baryons of the ectocolor gauge group with masses much less than the ectocolor confinement scale $\Lambda_E$. Identifying the ectobaryons as our dark matter allows us to circumvent a major hurdle of thermal production of confined baryonic dark matter in models with fermions in complex representations. In such models, both the dark matter mass ($\sim \Lambda_E$) and cross section ($\sim \Lambda_E^{-2}$) are set by the same confinement scale, and so obtaining the correct abundance forces the dark matter to be extremely heavy, $\sim 20~$TeV, at the edge of the unitarity limit \cite{Griest:1989wd}. 

In our model, the dark matter mass is proportional to the fundamental fermion masses (as is the case with the pions of $SU(3)_C$), allowing for richer phenomenology in both the early Universe and today. In particular, while many models of confined dark matter with masses $\ll 20$~TeV are forced to rely on asymmetric production mechanisms (see {\it e.g.}~technicolor/composite Higgs \cite{Nussinov:1985xr,Gudnason:2006ug,Hur:2007uz,Hur:2011sv,Ryttov:2008xe,DelNobile:2011je} and quirky \cite{Kribs:2009fy} dark matter), ectocolor dark matter is thermal, and so is composed of ectobaryons and their antiparticles. This allows for indirect detection signals, without requiring small ectobaryon-number violating terms \cite{Buckley:2011ye,Tulin:2012re}. Furthermore, as the ectobaryons are themselves only in thermal equilibrium with the unstable ectomesons, the freeze-out process in the early Universe is more complicated than the standard solutions to the Boltzmann equation, potentially resulting in thermal dark matter with present-day cross sections significantly lower than the naive expectation.

In this paper, we first describe the general formalism of ectocolor dark matter and the freeze-out process in the early Universe. We then describe the possible direct and indirect detection signals for these models, though these are small for the minimal model. We conclude with some of the unique collider phenomenology that can be realized in ectocolor dark matter.

\section{Ectocolor Dark Matter}

The goal of this paper is to introduce a viable thermal dark matter candidate which is composed of fundamental fermions bound into a composite object by a confining gauge force, with a mass much less than the unitary limit of dark matter ($\sim 20$~TeV). In particular, we will be interested in models in which the fundamental fermions carry electroweak charge, but form a stable composite state which is neutral, leading to interesting and viable phenomenology.  An obvious starting point would be a QCD-like theory, with the equivalent of the neutron as the dark matter candidate; stability of the neutron equivalent can be easily arranged by assignment of the ``quark" masses, $m_d \ll m_u$.

A strong constraint in building a QCD-like model of dark matter is that the stable baryons have a mass set by the same scale that sets their interaction cross section.  As a non-perturbative force, we cannot derive precise results without the lattice, but for our purposes, an estimate is sufficient. Roughly, the lightest baryon mass is set by the confinement scale $\Lambda$, at which the gauge coupling is driven non-perturbative by renormalization group evolution (for QCD, $\Lambda \sim 1$~GeV). The resulting baryon has a self-scattering cross section essentially given in the low-momentum transfer limit by a black disk approximation, with a physical size set by the same scale, $\sigma \sim \Lambda^{-2}$. While low velocity effects might greatly increase this cross section, it is difficult to see how it could be reduced.  In QCD this naive estimate would give a cross section for low-energy nucleon-nucleon annihliation of $\sigma_{N\bar{N}} \sim 0.4$ mb, whereas experimentally-measured cross sections are $\mathcal{O}(100~\mbox{mb})$ \cite{Armstrong:1987nu}.

In order to determine the relic abundance of a particle in thermal equilibrium with the bath of Standard Model particles in the early Universe, we must solve the complete Boltzmann equation. Later in this section, we will go into more detail, but for the moment, it suffices to note that, for a standard freeze-out calculation, the observed dark matter abundance is obtained when the velocity-independent ($s$-wave) cross section is $\sigma \sim 1$~pb. Using the simple black disk approximation, this translates into a confinement scale, and thus a dark matter particle mass, of $\Lambda \sim 20$~TeV. This is approximately the same as the maximum dark matter mass allowed by unitarity arguments  \cite{Griest:1989wd}, which is in retrospect not surprising.

Therefore, if we are to obtain a dark matter candidate out of strongly coupled physics with the mass as a free parameter, we must either turn to non-thermal production mechanisms \cite{Nussinov:1985xr,Barr:1990ca,Barr:1991qn,Gudnason:2006ug,Kribs:2009fy,DelNobile:2011je}, or find some way to divorce the annihilation cross section from the mass. In pursuing the latter, we again build our intuition from QCD. While baryons have mass $\sim \Lambda$, the pions are significantly lighter. This is because they are pseudo-Nambu-Goldstone bosons (PNGBs) of an approximate flavor symmetry. In the absence of quark masses, they would themselves be massless. The interactions between the pions and the other strongly interacting bound states is set by a parameter $F_\pi$, which is itself set by the confinement scale $4\pi F_\pi \sim \Lambda$ through non-perturbative physics.

However, while the pions of QCD have the desired relation between interaction strength and mass, it is non-trivial to create a model in which they are stable on cosmological timescales.  This is because pions are fundamentally composed of a fermion and anti-fermion pair, leading to many possibilities for self-annihilation and thus pion decay.  (It is possible to construct a dark matter model with stable pions, by use of a discrete symmetry similar to $G$-parity \cite{Bai:2010qg}, although a Peccei-Quinn symmetry in the new sector must be invoked in order to forbid higher-dimension operators which would violate the discrete symmetry.)  Stated in this way, the solution to our model-building problem is clear: we want our dark matter candidate to be a PNGB of an approximate flavor symmetry and {\it also} a gauge singlet combination of fundamental fermions, rather than fermion-antifermion pairs.

This cannot be achieved in models where the fermions are charged under complex representations of the strong gauge group; again using QCD as our example, since quarks are in {\bf 3} and anti-quarks in $\mathbf{\bar{3}}$, we cannot build singlets out of quark pairs only. However, if the gauge group has fermions in real or pseudo-real representations, then new composite operators are possible. The most familiar example (and the one we will use for explicit calculation in this paper) is $SU(2)$, the fundamental representation ${\bf 2}$ of which has the property that ${\bf \bar{2}} = {\bf 2}$ (up to a symmetry rotation). In such models, the spontaneous breaking of chiral symmetry will yield two types of PNGBs: mesons composed of quark-antiquark pairs, which decay; and baryons of quark-quark pairs, which can be stabilized by an analogue of baryon number.

For the remainder of this paper, we specialize the dark matter confining gauge group to $SU(2)_E$ ($E$ for ectocolor). Other choices are possible, such as $SO(N)$ or $Sp(N)$ gauge groups or fermions in adjoint representations of $SU(N)$, and may add additional complications to the cosmology and collider phenomenology.  However, our simple model captures the salient features. The confinement scale of $SU(2)_E$ is $\Lambda_E$; as we will show, to have dark matter with masses of ${\cal O}(100~\mbox{GeV})$, $\Lambda_E$ will generally be on the order of a few TeV.

The particle content of our benchmark model is shown in Table~\ref{tab:content}. The light fundamental fermions consists of two ectoquarks, $Q_u$ and $Q_d$ (up- and down-type), with opposite electric charges. Unlike in technicolor models, we assign only vector-like charges to the ectoquarks, and in our minimal scenario do not give $SU(2)_L$ charges. We impose a global $U(1)_X$ symmetry on the ectoquarks, which results in a conserved ``ectobaryon number'' (equivalently, dark matter number). As a result, the Lagrangian is
\begin{equation}
{\cal L} \supseteq i\bar{Q}_u \slashed{D} Q_u + i\bar{Q}_d \slashed{D} Q_d + m_u \bar{Q}_u Q_u + m_d \bar{Q}_d Q_d, \label{eq:lagrangian} 
\end{equation}
with $m_u$ and $m_d$ free parameters. By assumption $\Lambda_E \gg m_u, m_d > 0$, with $m_u \sim m_d \equiv m_q$, leading to an approximate global symmetry in which the $Q_u$, $Q_d$, $\bar{Q}_u$ and $\bar{Q}_d$ fields can be rotated into each other. Additional ectoquarks could be present in the full theory, but we assume that they are heavy enough that there is no approximate flavor symmetry (again, this constraint can be relaxed, and leads to a more complicated PNGB sector).

\begin{table}[ht]

\begin{tabular}{|c||c|c|c|c|c|}
\hline
 & $SU(2)_E$ & $SU(3)_C$ & $SU(2)_L$ & $U(1)_Y$ & $U(1)_X$ \\ \hline \hline
$Q_u$ & $\bf{2}$ & $\bf{1}$ & $\bf{1}$ & $+1/2$ & $+1/2$ \\ \hline 
$Q_d$ & $\bf{2}$ & $\bf{1}$ & $\bf{1}$ & $-1/2$ & $+1/2$ \\ \hline 
 \end{tabular}

\caption{Particle content and charges of the ectoquarks in our minimal model. \label{tab:content}}
\end{table}

In QCD, the light quark sector of $N_f$ flavors contains a $SU(N_f)_L\times SU(N_f)_R$ approximate global symmetry. When the $SU(3)_C$ gauge coupling becomes non-perturbative, the quark-anti-quark vacuum expectation value becomes non-zero: for small $N_f$, $\langle \bar{q}q \rangle \sim \Lambda_{\rm QCD}^3$. For QCD, with the quarks in complex (triplet) representations of $SU(3)_C$, this vev leads to the breaking $SU(N_f)_L \times SU(N_f)_R\to SU(N_f)_V$. For the two light quarks, the resulting three broken generators become the pion PNGBs. 

In a ectocolor model with ectoquarks in real or pseudoreal representations, the fields $Q$ and $\bar{Q}$ exist in the same representation, and additional global rotations are preserved.  As a result, the chiral symmetry group is enhanced from $SU(N_f) \times SU(N_f)$ to $SU(2N_f)$.  For $SU(2)_E$, the field redefinition
%%%%%%%%%%
\beqs
\psi_{i,L} &\equiv& -i \bar{Q}_{i,L} \sigma_2 \tau_2 ,\\
\bar{\psi}_{i,L} &\equiv& i \sigma_2 \tau_2 Q_{i,R}, \nonumber
\eeqs
%%%%%%%%%%
makes the enhanced symmetry manifest in the Lagrangian.  Here $\sigma_2$ and $\tau_2$ are the second Pauli matrix acting in spin and ectocolor space, respectively, and $i=u,d$ are the flavor indices.

For the pseudoreal representations which we focus on, the resulting breaking induced by the non-perturbative physics at $\Lambda_E$ is 
\begin{equation}
SU(2N_f) \to Sp(2N_f). \label{eq:breakingsym}
\end{equation}
As $SU(2N_f)$ has $4N_f^2-1$ generators and $Sp(2N_f)$ has $2N_f^2+N_f$, for our minimal model ($N_f=2$) there are $15-10=5$ broken generators, and so five PNGB fields. These fields can be broken down to three mesons without $U(1)_X$ number and two neutral baryons with $U(1)_X = \pm 1$:
\begin{eqnarray}
& \Pi^+ = Q_u \bar{Q}_d,~\Pi^- = Q_d\bar{Q}_u,~\Pi^0=\frac{1}{\sqrt{2}}\left(Q_u\bar{Q}_u - \bar{Q}_d\bar{Q}_d\right) & \\
& N = Q_uQ_d,~\bar{N} = \bar{Q}_u\bar{Q}_d. &
\end{eqnarray}
The fields $N$ and $\bar{N}$ will be our dark matter.

If the ectoquark masses were zero, then the $\Pi$ and $N$ fields would be exact Nambu-Goldstone bosons and hence massless. Assuming a common mass term $m_u = m_d \equiv m_q \ll \Lambda_E$, all three fields would have a common mass $M$ at tree level, related to the confinement scale by
\begin{equation}
F_\Pi^2 M^2 = m_q \langle \bar{Q} Q \rangle \simeq m_q \Lambda_E^3. \label{eq:massF}
\end{equation} 
Here, $F_\Pi$ is the ectocolor pion decay constant. Its value must be extracted from the non-perturbative physics, either from measurement or by lattice calcuation, but we can make the approximation (true in QCD) that
\begin{equation}
4\pi F_\Pi \sim \Lambda_E.
\end{equation}
Therefore, if $M$ is to be on the order of, say $200$~GeV, then 
\begin{equation}
m_q \sim 30~\mbox{MeV}\times \left(\frac{M}{200~\mbox{GeV}}\right)^2\left(\frac{700~\mbox{GeV}}{F_\Pi}\right).
\end{equation}

The charged meson $\Pi^\pm$ will gain an electromagnetic loop correction, raising its mass above that of the $N$ and $\Pi^0$. We may estimate this mass splitting as $\Delta M^2 \sim \alpha M^2/4 \pi$, lifting the charged meson's mass by $\sim 2\%$ above the neutral states.  Introduction of an ``isospin" splitting $\delta m = (m_u - m_d) \neq 0$ does not shift any of the PNGB masses at leading order \cite{Gasser:1983yg}, so we will assume $\delta m = 0$ for this work. Due to the small splittings between the charged and neutral states, the LEP-II bounds on new charged particles limits $M\gtrsim 90$~GeV \cite{Beringer:1900zz}, as we will discuss in more detail in Section~\ref{sec:collider}.

\subsection{Early-Universe Interactions and Decays} \label{sec:interactions}

To determine the phenomenology of these ectocolor singlets, in particular the relic abundance after thermal freeze-out, we must calculate their self-interactions and interactions with the Standard Model at energies much below $\Lambda_E$, where sensitivity to the internal ectoquark structure is suppressed by $F_\Pi$.  Since we are working with pseudo-NGB states which are light relative to the strong-coupling scale, we can work in the effective framework of chiral perturbation theory ($\chi$PT), expanding in the interaction momentum $p^2 / \Lambda_E^2 = p^2 / (4\pi F_\Pi)^2$.  This expansion will work well both in the early Universe at temperatures $T \ll 4\pi F_\Pi$, and for decays and self-interactions of cold dark matter in the present Universe.  We also assume $M \ll 4\pi F_\Pi$, since the convergence of $\chi$PT requires the violations of chiral symmetry induced by the ectoquark masses to be relatively small.

Chiral perturbation theory is an effective field theory, whose parameters are determined by the dynamics of the underlying strongly-coupled gauge theory.  In general, these parameters are very poorly known for theories other than QCD.  With the enhanced symmetry arising from real or pseudo-real fermions, the form of the chiral Lagrangian is changed somewhat \cite{Bijnens:2009qm}, but it is still qualitatively similar to the familiar structure arising from QCD.

We begin with the PNGB self-scattering. As the $N$ and $\Pi^0$ fields are electrically (and color) neutral, this is the only interaction which can keep them in equilibrium with other fields (in particular, the $\Pi^\pm$) when the temperature is $\ll \Lambda_E$. At leading order in the chiral expansion, the scattering cross section of any two PNGBs $P = \{\Pi^\pm,\Pi^0,N,\bar{N}\}$ with center of mass energy $\sqrt{s}$ is \cite{Weinberg:1966kf,Gasser:1983yg,Bijnens:2011fm}   
%%%%%%%%%%
\beq
\sigma(P_1P_2\to P_3P_4) = \frac{M^2}{16\pi F_\Pi^4} \frac{(s/M^2-1)^2}{s/M^2}. \label{eq:Pscattering}
%\langle\sigma v\rangle_N \equiv \langle \sigma v\rangle(P_1P_2\leftrightarrow P_3P_4) & = & \frac{M^2}{4\pi F_\Pi^4}\left( \frac{9}{16}+\frac{15}{16}v^2\right). \label{eq:Pscattering}
\eeq
%%%%%%%%%%
Because of the residual chiral symmetry making all of the PNGB masses degenerate, this is a process occurring at kinematic threshold, which is unusual for an inelastic process relevant for studying the thermal history of the Universe.  In particular, since the cross section at low energies is independent of the incoming particle velocity, we have that $(\sigma v) \propto v$, which complicates the derivation of the thermal average.  We make use of a general result for the calculation of $\langle \sigma v \rangle$ from the cross section in an inelastic $2 \rightarrow 2$ process \cite{Gondolo:1990dk}:
%%%%%%%%%%
\beq
\langle \sigma v \rangle = \frac{1}{8M^4 T K_2(M/T)^2} \int_{4M^2}^\infty \sigma \sqrt{s} K_1(\sqrt{s}/T) (s-4M^2)  ds, \label{eq:sigmav}
\eeq
%%%%%%%%%%
where $K_n$ are Bessel functions of the second kind, $v$ is the M\o ller velocity as defined in Ref.~\cite{Gondolo:1990dk}, and the thermal average is taken over Maxwell-Boltzmann thermal distributions at temperature $T$.  The analytic result of this integration for the cross section Eq.~\eqref{eq:Pscattering} cannot be expressed in simple terms, so we carry out the integration numerically for our study of the thermal abundance to follow.  However, the result can be expanded in the limit of large $x \equiv M/T$, yielding
%%%%%%%%%%
\beq
\left. \langle \sigma v \rangle (P_1P_2\to P_3 P_4)\right|_{T \ll M} = \frac{M^2}{\pi^{3/2} F_\Pi^4} \left(\frac{9}{16x^{1/2}} + \frac{255}{256x^{3/2}} + \mathcal{O}(x^{-5/2}) \right).
\eeq
%%%%%%%%%%
Unusually, the leading scaling of the thermally-averaged cross section is $T^{1/2}$, rather than $T^0$.  This will lead to a suppression of the annihilation rate in the current Universe, when compared to the early-Universe rate. Essentially, the velocity-averaged cross section is ``half-way'' between $s$-wave ($\propto v^0$) and $p$-wave ($\propto v^2$) processes. We will return to this point when we consider indirect detection signals in Section~\ref{sec:indirect}.

As a neutral particle under the Standard Model gauge groups, the $N$ field cannot be directly in thermal equilibrium with the bath of Standard Model fields (barring operators suppressed by powers of $\Lambda_E$ which are not relevant when temperatures are at or below $M$. See Section~\ref{sec:direct}). Instead, the interaction of Eq.~\eqref{eq:Pscattering} keeps the baryons in  equilibrium with $\Pi^0$ and $\Pi^\pm$, and the electromagnetic interaction of the $\Pi^\pm$ keeps that field in equilibrium with the bath. In the kinematic regime of interest, the velocity averaged thermal cross section of this interaction is
%%%%%%%%%%
\beq
\langle \sigma v\rangle_{\rm e.m.} \equiv \langle  \sigma v\rangle(\Pi^+\Pi^- \leftrightarrow \gamma\gamma)+\langle  \sigma v\rangle(\Pi^+\Pi^- \leftrightarrow f\bar{f}) \approx \frac{12\pi\alpha^2}{M^2} + \mathcal{O}(v^2) \label{eq:photonscattering}
\eeq
%%%%%%%%%%
Here, $f$ are Standard Model fermions with mass $m_f$ and $N_c$ colors, and we have assumed that $m_F \ll M$. Again using the $M=200$~GeV benchmark, the $s$-wave component of the cross section into photons is $\sim 20$~pb. While this is much greater than the canonical cross section for dark matter, it still implies that the $\Pi^\pm$ must decay, otherwise they would constitute a significant fraction of the Universe's matter density after thermal freezeout.  We ignore velocity-dependent corrections, which will not matter in the thermal history of our model.
 
The decay of the charged mesons must proceed through additional high scale physics, as no particle in Table~\ref{tab:content} couples to the $W^\pm$. One possibility is to add new heavy ectoquarks that are doublets of $SU(2)_L$ (either in vector or chiral representations). As long as their masses $m_Q$ are $\gg m_q$, they will not be part of the approximate flavor symmetry which leads to the light PNGB quintuplet. Therefore, after $SU(2)_E$ becomes non-perturbative the minimum mass for the bound states containing these heavy quarks is the confinement scale $\Lambda_E$ (though they can be heavier, if $m_Q\gtrsim \Lambda_E$).

In this scenario, the decay of the light $\Pi^\pm \to (W^\pm)^* \to f\bar{f}'$ proceeds through a loop of strongly coupled bound states containing these $SU(2)_L$-charged ectoquarks. This loop factor leads to a suppression of the coupling to the $W$ boson by a factor of $(M/4\pi F_\Pi)^2$. Interestingly, because the $\Pi^\pm$ is a scalar decaying through the chirally coupled weak force, there must be a spin-flip in order to conserve angular momentum. This results in a preference to decay into the heaviest Standard Model weak doublet that is kinematically available, with a width of 
\begin{equation}
\Gamma(\Pi^\pm \to W^* \to f\bar{f}') = N_c\frac{G^2_F m_f^2 M (M^2-m_f^2)^2}{2^{10} \pi^5 F_\Pi^2} \left(\frac{m_W^2}{M^2-m_W^2} \right)^2, \label{eq:pidecaySU2}
\end{equation}
here $N_c$ is the color factor of the Standard Model particles (3 for quarks, 1 for leptons). Again using our benchmark numbers, this leads to a decay of a 200~GeV $\Pi^\pm$ into top-bottom quark pairs with a width of $3.8 \times 10^{-8}$~GeV. For $M < m_{\rm top}$, the decay prefers $\tau/\nu_\tau$ and charm-strange pairs, with approximate branching ratios of $0.6$ and $0.4$ respectively.

An alternative possibility that can lead to $\Pi^\pm$ decay is that both the ectoquarks and (some) Standard Model fields are charged under a new gauge group with a $W'$, allowing direct coupling between the ectoquarks and the Standard Model. The most obvious possibilities for such a new gauge group are another $SU(2)_L$ coupling to the Standard Model left-handed quark and lepton doublets, or a $SU(2)_R$ group coupling with the right-handed quark and lepton doublets (including a right-handed neutrino) \cite{Mohapatra:1974hk,Senjanovic:1975rk}. 

While other gauge groups might be found, both these options share the preferential decay into the heaviest kinematically allowed fermion pair that was found in the ectohadron-mediated decay Eq.~\eqref{eq:pidecaySU2}. If the new gauge group has the same coupling strength $g$ as $SU(2)_L$, we can parametrize the coupling strength by $G_F' = G_F \left(m_W/m_{W'}\right)^2$. Depending on the assumptions placed on the flavor structure of the $W'$ model, the current collider bounds limit the $W'$ mass to be above a few TeV. The most stringent bounds come from the $W' \to \ell\nu$ channels; here $m_{W'} > 2.5$~TeV, assuming a Standard Model gauge coupling \cite{Aad:2011yg,Chatrchyan:2012meb}. Similar bounds can be set by low energy observables, see Ref.~\cite{Beringer:1900zz} for a review. From this, we can estimate the $\Pi^\pm$ width when mediated by a $W'$ (assuming $m_{W'} \gg M$):
\begin{equation}
\Gamma(\Pi^\pm \to W'^* \to f\bar{f}') = N_c\frac{G'^2_F m_f^2 F_\Pi^2 (M^2-m_f^2)^2}{4 \pi M^3}, \label{eq:pidecayWp}
\end{equation}
which, for $m_{W'} = 3$~TeV, $M=200$, $F_\Pi = 700$~GeV, gives a width of $2\times 10^{-6}$~GeV into top-bottom pairs.

Finally, we must consider the decay of the neutral meson, $\Pi^0$. Unlike the charged meson, no additional physics is needed to allow this particle to decay. Just as with the neutral pion of QCD, the two charged constituents inside the $\Pi^0$ will allow annihilation directly into gauge bosons. However, unlike the $\pi^0$ in QCD, decays to two gauge bosons does not necessarily dominate; for a wide range of parameter space, nearly $100\%$ of decays will go to SM fermion pairs. 

We start with the decay to two gauge bosons. Since in our minimal model the $Q_u$ and $Q_d$ only have $U(1)_Y$ hypercharge, the annihilation will proceed into photons and $Z$ bosons. Critically, with only two ectoquarks with equal and opposite charges, there is no contribution from the axial anomaly, and so the decay is suppressed by an additional factor of $(M/4\pi F_\Pi)^2$ compared to the equivalent rate for QCD pions. Therefore,
\begin{eqnarray}
\Gamma(\Pi^0 \to \gamma\gamma) & = & \left(\frac{\alpha}{\pi F_\Pi} \frac{M^2}{16\pi^2F_\Pi^2} \right)^2\frac{M^3}{64\pi} = \frac{\alpha^2 M^7}{2^{14} \pi^7 F_\Pi^6} \label{eq:gammapi0gg} \\
\Gamma(\Pi^0 \to \gamma Z) & = & \frac{\alpha^2 \tan^2\theta_WM^7}{2^{13} \pi^7 F_\Pi^6} \left(1-\frac{m_Z^2}{M}\right),  \label{eq:gammapi0gZ} \\
\Gamma(\Pi^0 \to Z Z) & = & \frac{\alpha^2 \tan^4\theta_WM^7}{2^{14} \pi^7 F_\Pi^6}\left(1-\frac{4m_Z^2}{M}\right)^{1/2}.  \label{eq:gammapi0ZZ}
\end{eqnarray}
In Eqs.~\eqref{eq:gammapi0gZ} and \eqref{eq:gammapi0ZZ}, we have assumed that $M$ is greater than the mass of $m_Z$ and $2m_Z$, respectively. For our benchmark this leads to a width of $3\times 10^{-13}$~GeV.  The decay amplitude into pairs of $W$ bosons vanishes at tree level.

\begin{figure}[ht]
\includegraphics[width=0.5\columnwidth]{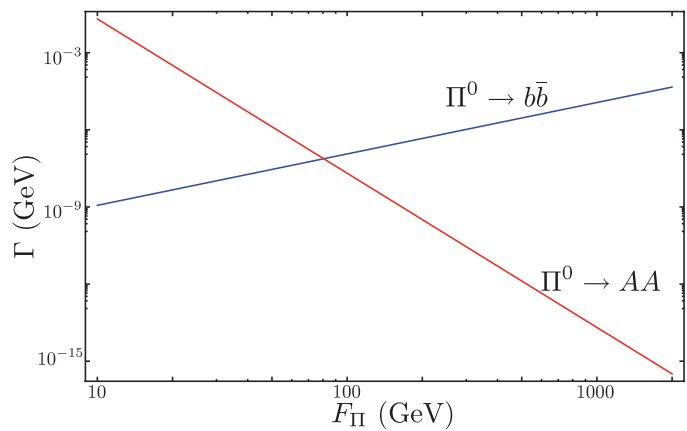}
\caption{Width $\Gamma$ of $\Pi^0$ decaying into gauge boson pairs $AA=\gamma\gamma,\gamma Z, ZZ$ (Eqs.~\eqref{eq:gammapi0gg}-\eqref{eq:gammapi0ZZ}) and into $b\bar{b}$ pairs (Eqs.~\eqref{eq:gammapi0bb}), as a function of $F_\Pi$. $\Pi^0$ mass $M$ is kept fixed at $M=200$~GeV.   \label{fig:Pi0decay}}
\end{figure}

In addition, there is a decay through a virtual $Z$ to Standard Model fermion pairs (decay through an off-shell photon is forbidden since the initial state is spin-0.)  In the Standard Model $\pi^0$, this mode is highly suppressed (BR $\sim 10^{-8}$ to $e^-e^+$ pairs). However, due to the wide range of $M$ and $F_\Pi$ available, and the additional loop suppression inherent in Eqs.~\eqref{eq:gammapi0gg}-\eqref{eq:gammapi0ZZ}, generically we expect this decay channel to completely dominate the $\Pi^0$ decay. As with the $W$- or $W'$-mediated decay of the $\Pi^\pm$, this decay mode of the $\Pi^0$ requires a spin-flip of the SM fermion, and so will couple to the heaviest state kinematically available (bottom quarks for $M<2m_{\rm top}$, tops otherwise). The width is given by
\begin{equation}
\Gamma (\Pi^0 \to Z^* \to f\bar{f}) = N_c \frac{G_F^2 \sin^4\theta_W Q_Z^2 F_\Pi^2 m_f^2 (M^2-m_f^2)}{8\pi M} \left(\frac{m_Z^2}{M^2-m_Z^2}\right)^2, \label{eq:gammapi0bb}
\end{equation}
where $Q_Z = T_3^f-Q_f \sin^2\theta_W$ is the $Z$-coupling of the SM fermion $f$. For our benchmark mass point, the decay into bottom quarks has a width of $1\times 10^{-5}$~GeV, and so is completely dominant over the two-gauge boson decays. In Fig.~\ref{fig:Pi0decay}, we show the two widths as a function of $F_\Pi$ for a fixed $M = 200$~GeV. As can be seen, only at very small $F_\Pi$ does the two-gauge boson mode dominate. However, this is precisely the region of parameter space where our $\chi$PT expansion is untrustworthy.

\subsection{Thermal History}

We now have all the requisite pieces to calculate the early Universe history of ectocolor dark matter. In principle, the story is relatively straightforward:  above $\Lambda_E$, the ectoplasma is kept in thermal equilibrium through the hypercharge interactions of the ectoquarks. After confinement, the $N/\bar{N}$ pairs are kept in thermal equilibrium with both the $\Pi^0$ and $\Pi^\pm$ fields by the hadronic scattering of Eqs.~\eqref{eq:Pscattering} and~\eqref{eq:sigmav} ($\langle \sigma v\rangle_{N}$). The $\Pi^\pm$ fields are in turn in equilibrium with the Standard Model bath, due to the electromagnetic interactions of Eq.~\eqref{eq:photonscattering} ($\langle \sigma v\rangle_{\rm e.m.}$). At the same time, $\Pi^0$ and $\Pi^\pm$ particles are decaying away, but this cannot deplete the overall number density as long as the particles are strongly coupled to their respective baths. The differential equations controlling this behavior are a set of three coupled Boltzmann equations:
\begin{eqnarray}
Y_{DM}'(x) & = & \frac{s(x) x}{H}\left[-\frac{\langle \sigma v\rangle_{N}}{2}(Y_{DM}^2(x)-Y_{\Pi^\pm}(x)^2)-\frac{\langle  \sigma v\rangle_{N}}{2}(Y_{DM}^2(x)-4Y_{\Pi^0}(x)^2) \right] \label{eq:Nboltzmann} \\
Y'_{\Pi^0}(x) & = & \frac{s(x) x}{H}\left[-\langle  \sigma v\rangle_{N}(Y_{\Pi^0}^2(x)-\frac{1}{4} Y_{DM}(x)^2)-\langle  \sigma v\rangle_{N}(Y_{\Pi^0}^2(x)-\frac{1}{4} Y_{\Pi^\pm}(x)^2) \right] -\frac{x\Gamma_{\Pi^0}}{H}\left[Y_{\Pi^0}(x)-Y_{\rm eq}(x)\right] \label{eq:pi0boltzmann} \\
Y'_{\Pi^\pm}(x) & = & \frac{s(x) x}{H}\left[-\frac{\langle  \sigma v\rangle_{N}}{2}(Y_{\Pi^\pm}^2(x)-Y_{DM}(x)^2)-\frac{\langle  \sigma v\rangle_{N}}{2}(Y_{\Pi^\pm}^2(x)-4Y_{\Pi^0}(x)^2)\right. \label{eq:piplusboltzmann} \\
 & & \left.-\frac{\langle \sigma v\rangle_{\rm e.m.}}{2}(Y^2_{\Pi^\pm}(x) - Y_{\rm eq}(x)^2) \right] -\frac{x\Gamma_{\Pi^\pm}}{H}\left[ Y_{\Pi^\pm}(x)-Y_{\rm eq}(x)\right]  \nonumber
\end{eqnarray}
Here, prime refers to differentiation with respect to $x \equiv m/T$, $H$ is the Hubble parameter, $s(x)$ is the entropy density, and the $Y$ functions are the particle number densities normalized by the entropy density. $Y_{\rm eq}$ is the equilibrium number density of the background bath. In reality there are five such Boltzmann equations, but the conditions $Y_N = Y_{\bar{N}} \equiv Y_{DM}/2$ and $Y_{\Pi^+} = Y_{\Pi^-} \equiv Y_{\Pi^\pm} / 2$ allow us to reduce to these three.

The presence of the decay terms $\Gamma_{\Pi^0}$ and $\Gamma_{\Pi^\pm}$ modifies the standard freeze-out cosmology somewhat. Without those terms, the charged ectomeson would freeze-out from the thermal bath at some $x_{\rm f.o.}$, when the rate of interactions mediated by $\langle \sigma v\rangle_{\rm e.m.}$ falls below the expansion rate of the Universe $H$. If $\langle \sigma v \rangle_N >  \langle \sigma v\rangle_{\rm e.m.}$, then the $\Pi^0$ and $N$ fields would still be bound to $Y_{\Pi^\pm}$, and depart from thermal equilibrium as their charged partner does. Since, at this point, all three annihilation channels ($N\bar{N}$, $\Pi^+\Pi^-$, and $\Pi^0\Pi^0$) are in equilibrium with each other, but only one combination ($\Pi^+\Pi^-$) can annihilate into the bath, the extra degrees of freedom pull the charged particles out of equilibrium earlier than one might expect, resulting in a larger relic abundance (by a factor of 3) than a single particle would possess.

If the cross-section inequality were reversed,  $\langle \sigma v \rangle_{\rm e.m.} >  \langle \sigma v\rangle_{N}$, then the $\Pi^0$ and $N$ particles would have already decoupled from the $\Pi^\pm$ when the latter decoupled from the bath. Therefore, if decays were negligible, the present-day relic abundance of $N$ would be set by the ectohadron interaction cross section, as this defines the time that the link connecting $N$ to the thermal bath is severed. 

However, in our model, the charged and neutral ectomesons will decay. If that decay occurs quickly enough, the particles will be unable to freeze-out from thermal equilibrium. Instead, the decay and reverse decay processes will cause the decaying species to track $Y_{\rm eq}$ past the point at which they would naively have departed from equilibrium. Roughly, this occurs when $\Gamma$ is large enough so that the time when decays are relevant ($x_{\rm decay}$) satisfies
\begin{equation}
x_{\rm decay} = \sqrt{\frac{H}{\Gamma}} < x_{\rm f.o.} \label{eq:decayconstraint}
\end{equation}
where $x_{\rm f.o.} \sim 25$ is the time at which freeze-out would occur if the particles were stable. For all reasonable values of $M$ and $F_\Pi$, this inequality is satisfied. Indeed, in order for decay to occur sufficiently late that freeze-out would occur, $\Gamma$ must be less than $\sim 10^{-17}$~GeV, corresponding to a particle lifetime $c\tau \sim 20$~m. At least for the charged $\Pi^\pm$ with masses $\lesssim 1$~TeV, such long lifetimes are experimentally ruled out by collider constraints (see Section~\ref{sec:collider}), even if the lifetime were unexpectedly large -- for example by a very high $W'$ scale in Eq.~\eqref{eq:pidecayWp}. For the neutral $\Pi^0$, Eq.~\eqref{eq:gammapi0bb} indicates that Eq.~\eqref{eq:decayconstraint} is satisfied for all $M$ and $F_\Pi$ of ${\cal O}(100-1000~\mbox{GeV})$.

Therefore, the $\Pi^\pm$ and $\Pi^0$ particles will remain in thermal equilibrium, and the $N/\bar{N}$ system will decouple at $x_{\rm f.o.}$, determined by the standard Boltzmann evolution of a particle in contact with a thermal bath with interaction cross section $2 \langle \sigma v\rangle_N$ (the factor of 2 accounts for the $N\bar{N} \leftrightarrow \Pi^+\Pi^-$ and $N\bar{N} \leftrightarrow \Pi^0\Pi^0$ channels); effectively this is co-annihilation \cite{Griest:1990kh} when all particles concerned have identical mass. It is interesting to note that the ectobaryon dark matter provides a natural way to allow the present-day annihilation cross section of the dark matter $N$ to differ by an integer factor from the cross section that controls freeze-out. The ratio between the two cross sections could presumably be increased in models with more light PNGBs, either by increasing the number of light fermion species charged under $SU(2)_E$ or by placing the ectoquarks in real or pseudoreal representations of larger Lie groups.

\begin{figure}[ht]
\includegraphics[width=0.6\columnwidth]{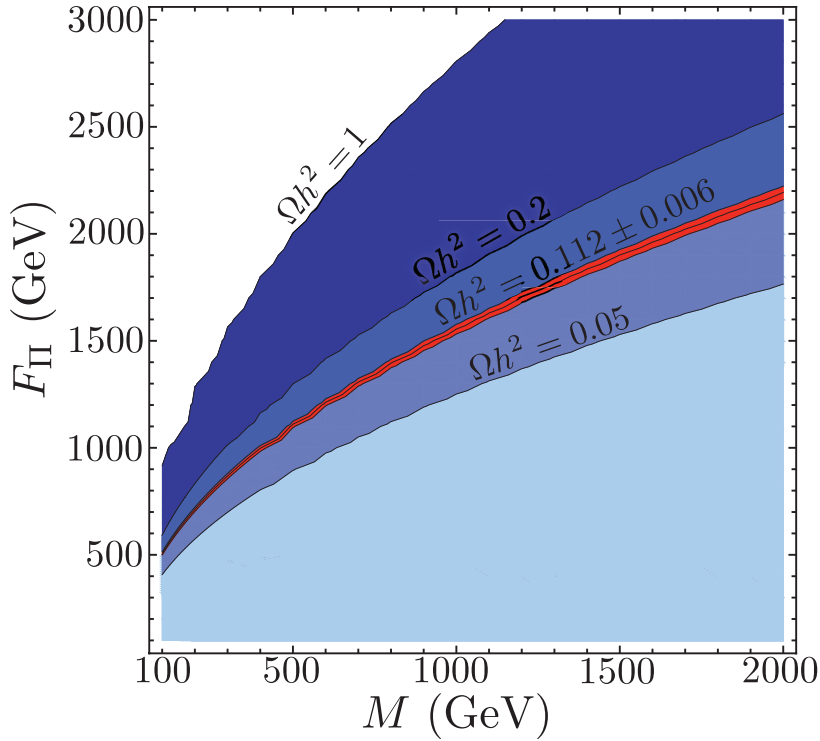}

\caption{Relic abundance $\Omega h^2$ of the $N/\bar{N}$ particles as a function of $M$ and $F_\Pi$, assuming $\Pi^\pm$ and $\Pi^0$ are prevented from going out of thermal equilibrium due to decays. The parameters providing the observed dark matter abundance $\Omega_{\rm DM}h^2 = 0.112\pm 0.006$~\cite{Beringer:1900zz} are shown in red.  \label{fig:decayOmega}}
\end{figure}

\begin{figure}[ht]
\includegraphics[width=0.6\columnwidth]{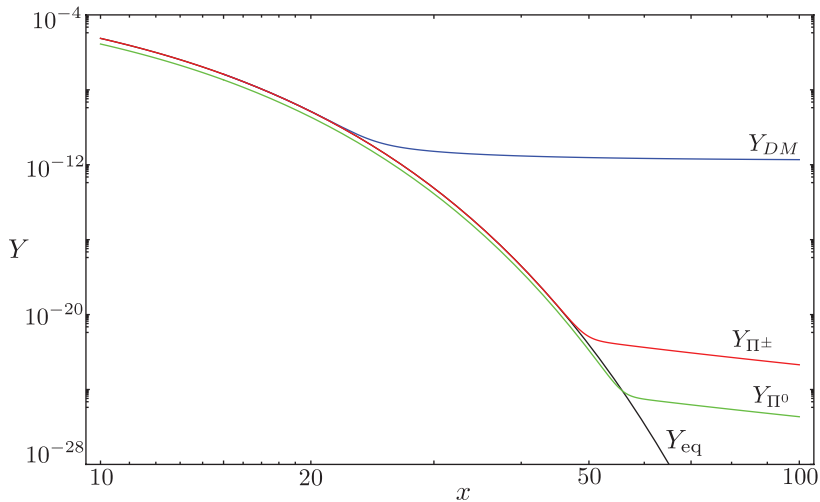}

\caption{Relic abundance evolution for $N/\bar{N}$ ($Y_{DM}$, blue), charged mesons $\Pi^\pm$ (red), and neutral mesons $\Pi^0$ (green) assuming $M=200$~GeV, $F_\Pi = 700$~GeV, and decay with widths given by Eq.~\eqref{eq:gammapi0bb} for $\Pi^0$ and Eq.~\eqref{eq:pidecaySU2} for $\Pi^\pm$. \label{fig:decayboltzmann}}
\end{figure}

Fig.~\ref{fig:decayOmega} shows the relic abundance of the $N/\bar{N}$ particles resulting from this set of assumptions. In Fig.~\ref{fig:decayboltzmann}, we show a sample relic abundance calculation for $M=200$~GeV and $F_\Pi=700$~GeV; at large $x$, the decaying particles depart from the thermal distribution due to the slow pair annihilation of $N \bar{N}\to \Pi~\Pi$. For this benchmark point, the relic abundance of the dark matter $N+\bar{N}$ is $\Omega h^2 = 0.105$, very close to the experimentally measured value of $ 0.112\pm 0.006$~\cite{Beringer:1900zz}. As can be seen from Fig.~\ref{fig:decayOmega}, this choice of parameters is not particularly fine-tuned. Though only a small region of $(M,F_\Pi)$ space gives the correct relic abundance, this is due to the precision of the experimental result, not to any required cancellation in the theory. Thus, we can say that ectocolor dark matter can provide a viable thermal candidate for dark matter over a wide range of parameter space.

\section{Experimental Signatures}

\subsection{Direct Detection}
\label{sec:direct}

Although our dark matter particle is an electroweak-neutral bound state, we expect it to interact with ordinary matter through photon and $Z$-boson exchanges with the bound ectoquarks.  In the present Universe, such interactions will occur only at very low energy, so they must be described in terms of electromagnetic form factors of the ectobaryons (we ignore weak exchanges, which will be further suppressed by $m_Z$ and $\sin^2 \theta_w$ in the context of direct detection.)

The interactions between composite dark matter particles and electromagnetic fields can be treated in an effective theory, expanding in the velocity $v^\mu$ of the dark matter.  The spin of the dark matter, in particular whether it is fermionic or bosonic, determines which operators will appear \cite{Bagnasco:1993st}.  For the model which we are considering in detail, the ectobaryon is a spin-0 boson.  The leading interactions will then proceed through the charge radius operator,
%%%%%%%%%%
\beq
\mathcal{L}_{\rm CR} \supset \frac{1}{\Lambda^2} \bar{N} N v_\nu \partial_\mu F^{\mu \nu},
\eeq
%%%%%%%%%%
and the EM polarizability,
%%%%%%%%%%
\beq
\mathcal{L}_{\rm pol.} \supset \frac{1}{\Lambda^3} \bar{N} N  F_{\mu \nu} F^{\mu \nu},~{\cal L}_{v-\rm pol.} \supset \frac{1}{\Lambda^3} \bar{N} N v_\mu v_\nu F^{\mu \sigma} F^\nu{}_\sigma.
\eeq
%%%%%%%%%%
In terms of the non-relativistic fields, the corresponding Hamiltonian for interaction of the ectonucleon with the EM field of an ordinary nucleus is given by \cite{Pospelov:2000bq}
%%%%%%%%%%
\beq
H = -\frac{e}{6}r_D^2 \frac{\partial}{\partial x_i} E_i - \frac{1}{2} (\chi_E E^2 + \chi_B B^2),
\eeq
%%%%%%%%%%
with resulting scattering cross-sections
%%%%%%%%%%
\beqs
\sigma_{R} & = & \frac{4\pi}{9} \mu_{ND}^2 Z^2 \alpha^2 r_D^4, \label{eq:sigmaR} \\
\sigma_{\chi} & \eqsim & \frac{144\pi}{25} \mu_{ND}^2 Z^4 \alpha^2 \frac{\chi_E^2}{r_0^2}, \label{eq:sigmachi}
\eeqs
%%%%%%%%%%
with $\mu_{ND}$ the reduced mass of the nucleus-dark matter system, and $r_0$ the charge radius of the target nucleus.  We will follow the choice of Ref.~\cite{Pospelov:2000bq} by taking $r_0 \sim (1.2\ \textrm{fm}) \sqrt[3]{A}$, with $A$ the mass number and $Z$ the atomic number of the target.  We neglect the interaction with the magnetic polarizability, which is generally sub-leading.

The coefficients $r_D$ and $\chi_E$ are given by low-momentum dynamics of the strongly-coupled dark sector, and non-perturbative techniques (such as lattice simulation) should be used in order to accurately determine them.  However, for $M \ll 4\pi F_\Pi$  we can reliably compute these quantities in the framework of chiral perturbation theory ($\chi$PT), although at higher orders these expressions will depend on unknown low-energy constants.  Many quantities have been computed to high order in $\chi$PT for QCD, but we cannot in general use these results directly, since the chiral Lagrangian for symmetry breaking with pseudo-real fermions must be modified to accommodate the enhanced symmetry group \cite{Bijnens:2009qm,Bijnens:2011fm}.  A one-loop computation of $r_D$ and $\chi_E$ within this modified framework would be quite interesting, but is beyond the scope of this work; here we will use symmetry arguments and the known QCD expressions in order to make order-of-magnitude estimates.

Because $Q_u$ and $Q_d$ carry equal and opposite electromagnetic charge in our model, in the limit $m_u \rightarrow m_d$ we find a $Z_2$ symmetry of the theory with respect to the field redefinition $Q_u \leftrightarrow Q_d$ and $A_\mu \rightarrow -A_\mu$ \cite{Kribs:2009fy}.  Since the electric field is odd under this symmetry, it is clear that the charge radius must vanish identically, $r_D^2 = 0$.  In the presence of a mass splitting $\delta m = m_u - m_d$, the $Z_2$ symmetry is broken, and we expect to generate a charge radius in some way parameterically small in $\delta m$.  For example, the charge radius of the $K^0$ in standard QCD $\chi$PT and at one loop is equal to \cite{Gasser:1984ux}
%%%%%%%%%%
\beq
\langle r_D^2\rangle_{K^0} = \frac{1}{16\pi^2 F^2} \log (M_K^2/M_\pi^2).
\eeq
%%%%%%%%%%
As expected, this expression vanishes in the limit $m_s \rightarrow m_d$.  Although the $K^0$ is similar to the ectonucleon with $\delta m \neq 0$ in that both are composite states of two equal-charge fermions with different masses, it is clear that the expression for the ectonucleon charge radius must be qualitatively different, since as we have noted, all of our Goldstone bosons have masses proportional to $(m_u + m_d)$.  At best the expression for $r_D^2$ for the ectonucleon will arise at one loop and be suppressed by $\delta m$ in some way, so we claim that as a conservative upper limit for small $\delta m$,
%%%%%%%%%%
\beq
\left. \langle r_D^2\rangle_{N}\right|_{\delta m \neq 0} \ll \frac{1}{16\pi^2 F_\Pi^2}.
\eeq
%%%%%%%%%%
We can convert this to an upper bound on the direct-detection cross section, using eq.~\ref{eq:sigmaR} and adjusting by the factor $\mu_{nD}^2 / (A^2 \mu_{ND}^2$) to convert to the standard ``WIMP-nucleon" cross-section.  We thus find
%%%%%%%%%%
\beq
\sigma_{SI}^{r_D} \ll (7.2 \times 10^{-49}~\mbox{cm}^2) \left(\frac{Z}{50}\right)^2 \left(\frac{130}{A}\right)^2 \left(\frac{700~\mbox{GeV}}{F_\Pi}\right)^4.
\eeq
%%%%%%%%%%

The electromagnetic polarizabilities can be obtained in $\chi$PT by examining the Compton scattering process $\gamma \pi \rightarrow \gamma \pi$.  For QCD, the leading contribution to polarizability of the $\pi^0$ occurs at $\mathcal{O}(p^4)$ and involves only the leading-order low-energy constants, since vertices of the form $\pi^0 \pi^0 \gamma$ and $\pi^0 \pi^0 \gamma \gamma$ are forbidden in the $\chi$PT Lagrangian through next-to-leading order \cite{Bijnens:1987dc}.  The argument given in the reference does not apply trivially  to the pseudoreal case, but it can be verified explicitly that the generators corresponding to the $N/\bar{N}$ states commute with the charge matrix $Q$, leading to the same result.

Again explicitly for QCD, the electric susceptibility of the $\pi^0$ is given at leading order by the expression \cite{Bellucci:1994eb}
%%%%%%%%%%
\beq
\chi_E^{(\pi^0)} = \frac{1}{96 \pi^2 F^2 m_\pi}.
\eeq
%%%%%%%%%%
Based on the arguments above, we expect the $N/\bar{N}$ polarizability to be the same at this order, up to $\mathcal{O}(1)$ factors.  We therefore use this formula to obtain a rough estimate for the direct-detection constraints on ectocolor dark matter through this operator.  Taking Eq.~\eqref{eq:sigmachi} and again adjusting by the factor $\mu_{nD}^2 / (A^2 \mu_{ND}^2$) to convert to the standard ``WIMP-nucleon" cross-section, we find
%%%%%%%%%%
\beq
\sigma_{SI}^{\chi_E} \approx (4.3 \times 10^{-52}\ \textrm{cm}^{2}) \left(\frac{Z}{50}\right)^4 \left(\frac{130}{A}\right)^{8/3} \left(\frac{200~\mbox{GeV}}{M}\right)^2 \left(\frac{700~\mbox{GeV}}{F_\Pi}\right)^4.
\eeq
%%%%%%%%%%

These cross-section estimates are relatively crude, and a more rigorous calculation in the framework of $\chi$PT, as well as lattice studies to fix the values of the low-energy constants in the chiral Lagrangian, will be needed for a precise understanding of direct detection in this model.  However, these estimates are quite far below existing experimental bounds, so we can at least be confident that our minimal construction is not constrained by direct detection currently.

\subsection{Indirect Detection}
\label{sec:indirect}

As the dark matter is composed of both ectobaryons $N$ and their antiparticles $\bar{N}$, self-annihilation can occur in the current Universe. However, the $N-\Pi$ scattering Eq.~\eqref{eq:Pscattering} provides the primary annihilation cross-section in the minimal model, with the resulting unstable ectomesons decaying into visible Standard Model particles. Since this cross section is proportional to velocity $v$, indirect signals are highly suppressed in the present day.

Dark matter in the Galaxy has $v \sim 10^{-3}c$. As the mass splitting between the charged and neutral states $M_{\Pi^\pm} - M \sim 100~$MeV is much larger than the available kinetic energy  $M v^2$, annihilation will primarily proceed through $N\bar{N} \to \Pi^0\Pi^0$. As discussed in Section~\ref{sec:interactions}, the $\Pi^0$ will quickly decay into the heaviest kinematically available fermion pairs ($b\bar{b}$ for $M< 2m_{\rm top}$, tops otherwise). 

Currently, the best bounds on dark matter annihilating into $b\bar{b}$ are obtained in the data set collected by the Fermi Gamma-Ray Space Telescope (FGST). In particular, bounds from the Galactic Center \cite{Hooper:2012sr} are the strongest, even for conservative assumptions on the dark matter profile in the Galaxy's inner region. We also include the somewhat weaker stacked dwarf galaxy limits \cite{GeringerSameth:2011iw,Ackermann:2011wa,Farnier:2011zz}. Note that, when applying indirect detection bounds, the $2\to4$ annihilation in our model contributes an overall factor of two which cancels with a factor of $1/2$, as our dark matter is not Majorana. 

\begin{figure}[ht]
\includegraphics[width=0.6\columnwidth]{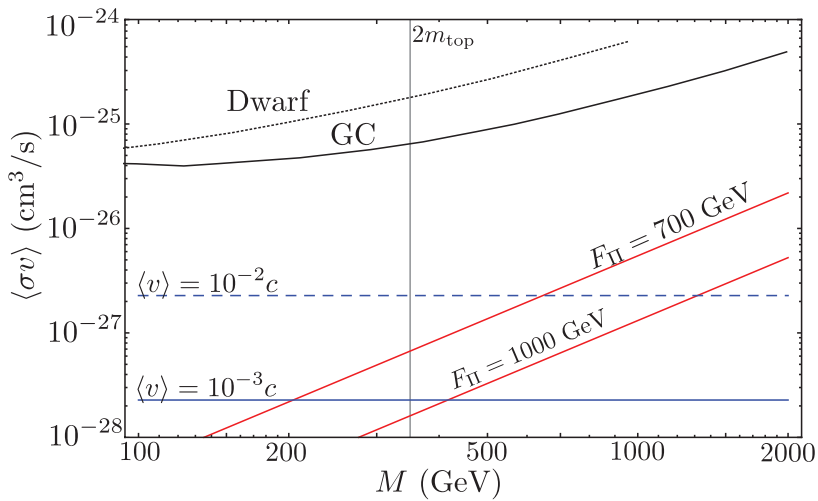}

\caption{Cross section for $N/\bar{N}$ annihilation in the Universe today as a function of $M$ for fixed $F_\Pi$ (red lines) as well as the $(M,F_\Pi)$ combinations that provide the correct relic abundance after thermal freeze-out assuming $\langle v\rangle = 10^{-3}c$ in the Galactic Center (blue line) or the unphysically large $10^{-2}c$ (dotted blue line). Also shown are the upper limits for annihilation in $b\bar{b}$ from the FGST dwarf stacking analysis \cite{GeringerSameth:2011iw,Ackermann:2011wa,Farnier:2011zz} and the Galactic Center assuming a NFW profile \cite{Hooper:2012sr}. Note that for $M>2m_{\rm top}$, we expect annihilation to proceed into top pairs, but we can safely extrapolate $b\bar{b}$ bounds to this region. \label{fig:indirectbounds}}
\end{figure}

The bounds from the Galactic Center and dwarfs have not been calculated in the $t\bar{t}$ channel. However, the resulting spectrum is not significantly different from the bounds on the $b\bar{b}$ channel \cite{Zaharijas:2006qb}, and so we are justified in extrapolating the bottom quark limits for the full range of mass $M$. In Fig.~\ref{fig:indirectbounds}, we show the upper limits on $\langle \sigma v\rangle_N$ followed by $\Pi^0$ decay into $b\bar{b}$ from both dwarf stacking and Galactic Center bounds, as well as the prediction for the values of $M$ and $F_\Pi$ which give the correct relic abundance. As can be seen, this places essentially no constraint on the $(M,F_\Pi)$ parameter space required for thermal dark matter in this minimal scenario.

\subsection{Collider Searches} \label{sec:collider}

The charged ectomesons $\Pi^\pm$ will be produced in pairs at colliders through Drell-Yan processes. For large $F_\Pi$, the internal structure of the ectohadrons is not relevant, and production will proceed as if the $\Pi^\pm$ were elementary particles. Though the small $F_\Pi$ regime can provide interesting and unique collider signatures (becoming `quirks' \cite{Kang:2008ea,Kribs:2009fy,Harnik:2011mv} as $F_\Pi \to 0$) we will leave such considerations to a later work, and assume that $F_\Pi$ is large enough so that the internal structure can be ignored.

The lack of observation of new charged particles in the stau channel at LEP-II allows us to extrapolate a fairly robust limit of $M\gtrsim 86.6$~GeV \cite{LEPSUSYWG:fk}, as $\Pi^\pm$ decays predominantly into $\tau^\pm \nu$ in this mass range, and so mimics the signal of $\tilde{\tau}$ pairs decaying into taus and massless neutralinos. 

\begin{figure}[ht]
\includegraphics[width=0.5\columnwidth]{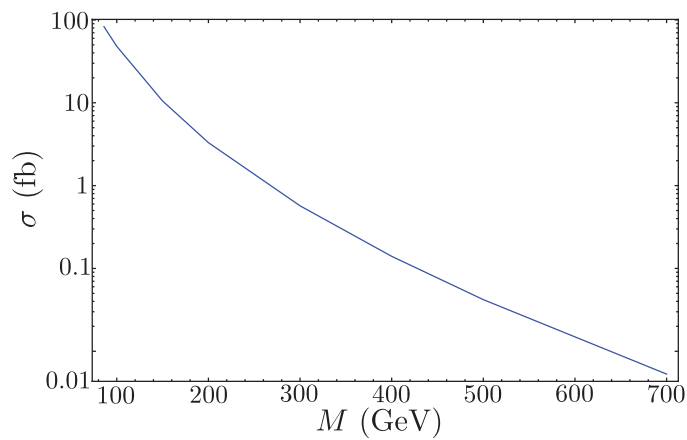}
\caption{$\Pi^+\Pi^-$ Drell-Yan production cross section at the LHC assuming $\sqrt{s} = 8$~TeV as a function of $\Pi^\pm$ mass $M$. \label{fig:LHCsigma} }
\end{figure}

At the LHC, the $\Pi^+\Pi^-$ production cross section depends only on the mass $M$, and is shown in Fig.~\ref{fig:LHCsigma}. For $M < m_{\rm top}+m_{b} = 176$~GeV, decays proceed through taus and neutrinos. Assuming this decay is prompt, the bounds from stau pair production followed by decays to taus and massless neutralinos are applicable. However, the current limits on such cross sections are ${\cal O}(3~\mbox{pb})$ \cite{CMS-PAS-SUS-11-007}, which does not place significant bounds on the ectomeson production. 

Above $\sim 176$~GeV, the $\Pi^\pm$ decay into top/bottom quark pairs. This is a somewhat unusual signature; while searches exist for single $tb$ resonances \cite{Aad:2012ej}, we are not aware of any search for $(t\bar{b})(\bar{t}b)$ final states. As the production cross section is small, dedicated searches for $\tilde{t} \to t \tilde{\chi}^0_1$ pair production \cite{CMS-PAS-SUS-11-007,ATLAS-CONF-2012-059,ATLAS-CONF-2012-070,ATLAS-CONF-2012-071,ATLAS-CONF-2012-073} are not sensitive to this production, even neglecting the difference in event acceptance due to the presence of two extra $b$-quarks. It is interesting to note that charged pseudoscalars are generically expected to decay into the heaviest fermion pair available, and so searches for $(t\bar{b})(\bar{t}b)$ final states might be relevant beyond this paper's particular dark matter model.

The previous discussion assumed the $\Pi^\pm$ decayed quickly on detector timescales. Roughly, this translates into widths greater than that of the $b$: $\Gamma \gtrsim 4\times 10^{-13}$~GeV.  If stable on detector timescales ({\it i.e.}~requiring $c\tau > 1$~m, $\Gamma_{\Pi^\pm} \lesssim 2 \times 10^{-16}~$GeV), constraints from the LHC on stable charged particles place a bound of $M\gtrsim 220$~GeV \cite{Chatrchyan:2012sp}. This bound only applies if the charged particles live long enough to pass through the calorimeters of CMS.  As can be seen from Eqs.~\eqref{eq:pidecaySU2} and \eqref{eq:pidecayWp}, this is far longer than the expected lifetime of the charged ectomesons, unless the decay is mediated by a new $W'$ with mass 
\begin{equation}
m_{W'} \gtrsim (250~\mbox{TeV})\left(\frac{F_\Pi}{700~\mbox{GeV}} \right)^2\left(\frac{M}{200~\mbox{GeV}}\right)\left(1-\frac{m_{\rm top}^2}{M^2}\right)^2,
\end{equation}
or via additional loops of heavy ectoquarks charged under $SU(2)_L$ with $F_\Pi \sim 10^8$~GeV. However, with a more modest $m_{W'}$ scale, the lifetime of the $\Pi^\pm$ could be on the scale of $\mu$m to cm. Such small displaced vertices would provide a very unique signature at the LHC, but it is unclear whether any existing analysis would be sensitive to the events. It should be noted that, as the decay would proceed through top and bottom quarks, the additional displacement of the $b$ decay makes such a search difficult. 

\section{Conclusion}

In this paper, we have introduced a new thermal dark matter candidate which is composite under a new confining force, which we dub ``ectocolor." Critically, we require the ectoquarks to be charged under real or pseudoreal representations of the ectocolor gauge group; the canonical example which we have considered in detail is the fundamental representation of $SU(2)$. The extra symmetry that this allows in the Lagrangian results in two types of pseudo-Nambu-Goldstone bosons of an approximate flavor symmetry: stable ectobaryons with a conserved quantum number, and unstable ectomesons. As the PNGBs have masses proportional to that of their constituent ectoquarks, the mass and couplings of the dark matter can be set independently of each other. This allows viable thermal dark matter from confining gauge groups with masses well below the unitary bound of $\sim 20$~TeV.

Ectocolor dark matter has a number of interesting features that distinguish it from a standard thermal candidate. As one of a number of PNGBs with degenerate masses, all of which are thermal equilibrium in the early Universe, the dark matter candidate $N$ essentially undergoes coannihilation with a potentially large number of particles. This allows the present-day self-interaction cross section of dark matter to appear significantly lower than the canonical value of 1~pb. In the explicit model we demonstrate in this paper, the ``coannihilation factor'' is 2, but this can easily be increased in models with a larger flavor symmetry.

Additionally, the dark matter annihilates into unstable ectomesons, which decay preferentially into the heaviest fermions kinematically available. This has potentially interesting predictions for the LHC, where the charged ectomesons can be pair-produced directly, though the production cross section is low.  The resulting $(t\bar{b})(t\bar{b})$ final states for some values of the parameters (possibly with displaced vertices) are an interesting -- and so far unexplored -- signature at the LHC.  This signature is furthermore a generic feature of charged pseudoscalar composites, which will generally decay with a mass flip in the final-state fermions.

The bounds from direct detection experiment on the model presented here are quite weak.  This is primarily due to a discrete symmetry of our minimal model under the interchange of the equal- and opposite-charged $Q_u$ and $Q_d$ fields, which eliminates the contribution from the charge radius operator.  It remains to be determined whether inclusion of a large mass splitting $m_u - m_d$, or constructing a more complicated model with additional charged states, can lead to more significant direct-detection signals.

The predicted rate of indirect detection is too low in the minimal model to provide a visible signal. However, if an annihilation channel was available that was not at a kinematic threshold, then the cross-section today would not be suppressed by $v$. Though loop annihilations into $\gamma\gamma$ in our minimal model are present, they are suppressed by several of orders of magnitude even from the $v$-dependent cross section. It is again perhaps useful to consider non-minimal models, where this would not be the case.

Interestingly, the minimal model does not provide a significant signal of $N\bar{N} \to \Pi^0\Pi^0 \to 4 \gamma$. This is somewhat surprising, as the Standard Model pion decays predominantly to photons. However, the combination of larger meson mass and fermion mass ($b$ instead of $e$) relative to the $Z$, as well as the lack of an axial anomaly greatly reduces the photon channel relative to the fermion final state. This makes it difficult to use this model to provide an explanation of the suggested 130~GeV line in the Fermi data \cite{Bringmann:2012vr, Weniger:2012tx} (see also Ref.~\cite{Tempel:2012ey,Su:2012ft,Su:2012zg}). Although confining dark models have been suggested as a source of the annihilation line \cite{Buckley:2012ws,Fan:2012sx}, a more complicated construction than our minimal scenario would be needed in order to explain this possible signal of dark matter as the result of a thermal ectocolor dark matter candidate.

Despite the fact that the dark sector is strongly coupled, the use of PNGB states as the dark matter makes analytic calculations using the framework of chiral perturbation theory quite tractable.  Explorations of the present model using $\chi$PT would be quite interesting, and will be necessary for e.g.~calculation of the direct-detection operators, and for exploration of the effects of mass splittings among the PNGBs.  The parameters of the chiral Lagrangian are determined by the underlying strong dynamics, and for the present example of $SU(2)$ the use of $1/N_c$ expansion is particularly unappealing, so lattice calculations of these low-energy constants may be an important input for more precise study of ectocolor models.
 
\begin{acknowledgments}
The authors wish to thank Scott Dodelson, Patrick Fox, Roni Harnik, Chris Hill, Michael Buchoff, and Dan Hooper for helpful advice and discussion. Fermilab is operated by Fermi Research Alliance, LLC, under Contract DE-AC02-07CH11359 with the United States Department of Energy.

\end{acknowledgments}
\bibliographystyle{apsrev}
\bibliography{hypercolor}

\end{document}